\documentclass[mathleft
]{an}
\usepackage{graphicx}
\usepackage{times}
\usepackage{tabularx}
\begin{document}

\Pagespan{789}{}
\Yearpublication{2006}%
\Yearsubmission{2005}%
\Month{11}%
\Volume{999}%
\Issue{88}%

\title{Identifications of FIRST radio sources in the NOAO Deep-Wide Field Survey}

\author{K.EL Bouchefry\inst{1}
\and  C.M.Cress\inst{2}\fnmsep\thanks{\email{cressc@ukzn.ac.za}\newline}\
}
\titlerunning{FIRST identifications}
\authorrunning{K. EL Bouchefry \& C.M.Cress}
\institute{
Astrophysics and Cosmology Research Unit, University of KwaZulu-Natal, Private Bag X01, Scottsville, 3209, South Africa}  

\received{30 May 2006}
\accepted{******}
\publonline{later}


\abstract{%
In this paper we present the results of an optical and near infrared
identification of $514$ radio sources from the FIRST survey (Faint Images of the
Radio Sky Survey at Twenty centimetres) with a
flux-density limit of $1$ mJy in the NOAO Deep-Wide Field Survey (NDWFS) Bo\"{o}tes field. Using optical ({\it{Bw,
    R, I}}) and {\it{K}} band data with approximate limits of {\it{Bw}}$\sim$
$25.5$ mag, {\it{R}}$\sim$$25.8$ mag, {\it{I}}$\sim$ $25.5$ mag and
{\it{K}}$\sim$$19.4$ mag, optical counterparts have been identified for $378$
of $514$ FIRST radio sources. This corresponds to an identification rate of
$34\%$ in four bands ({\it{BwRIK}}), $60\%$ in optical bands ({\it{BwRI}}) and
$74\%$ in {\it{I}} band. Photometric redshifts for these sources have been computed  using the \textit{hyperz} code. The inclusion of quasar template spectra in \textit{hyperz} is investigated. We note that the photometric data are, in many cases, best matched to templates with very short star-formation timescales and the inferred ages of identified galaxies depend strongly on the assumptions about the star-formation timescale. The redshifts obtained are fairly consistent with those expected from the K-z relation for brighter radio sources but there is more scatter in the K-z diagram at z$<$1. \\
}

\maketitle

\section{Introduction}
Radio sources detected in deep 1.4GHz surveys are generally drawn from two
populations (Condon, Cotton \& Broderick 2002). Brighter sources are mostly
active galactic nuclei (AGN) in which the radio emission is synchrotron
emission generated by energetic charged particles interacting with
magnetic fields. The particles are produced in the vicinity of the
supermassive black holes which reside at the centre of galaxies.
As one approaches 1$\,$mJy flux-densities, source populations include starforming galaxies in which radio emission is a combination of
synchrotron emission from charged particles, generated by star formation
activity, and thermal emission from hot gas in the environment of
starforming regions.

Understanding the sources found in 1.4GHz surveys is thus relevant to many areas of study including star formation histories and processes, accretion and black hole physics and astrophysical jet formation. Moreover, radio sources can be detected out to very high redshifts over large areas of sky and thus probe larger volumes than those typically observed in optical surveys. This makes them valuable cosmological probes. The clustering of these sources has been measured (Cress et al. 1996; Magliocchetti et al. 1998; Overzier et al. 2002; Blake, Ferreira \& Borrill 2004), as has the weak gravitational lensing (Chang et al. 2002) and these measurements can be used to put constraints on cosmology (Cress \& Kamionkowski 1998; Negrello, Magliocchetti \& De Zotti 2006; Kamionkowski et al. 1998), particularly when the redshift distribution is well known. In addition, if optical identifications allow sources to be separated into redshifts bins, then new information on the evolution of structure can be obtained.  

The largest previous sample of optical identifications for FIRST sources (Faint Images of the Radio Sky at Twenty centimeters; Becker, White \& Helfand 1995)
are based on the Cambridge Automated Plate-Measuring Machine (APM) survey (Maddox et al 1990). An identification program described by  
McMahon et al. (2002)
for $382,892$ sources in the north Galactic cap, using the APM scans
of POSS-I plates, was used as astrometric standard to improve the absolute
astrometry of the POSS plates: matching the radio and the optical catalogues
resulted in $70,000$ optical counterparts with E$<20$ and O$<21.5$ (where E and O bands are similar to R and B bands respectively). Magliocchetti et al. (2002) studied optical counterparts to FIRST sources in the 2-degree Field Galaxy Redshift Survey (2dFGRS, Colless et al 2001). Magliocchetti $\&$ Maddox (2002) present a detailed analysis of the properties and angular clustering of  $\sim 4000$
FIRST sources with bright optical counterparts in the APM survey. Ivezic at al. (2002) discuss the properties of $\sim 30,000$ sources
observed by the FIRST survey and the Sloan Digital Sky Survey (SDSS, York et al. 2000). The $ugriz$ observations of the SDSS have limiting AB magnitudes of 22.0, 22.2, 22.2, 21.3 and  20.5  respectively. SDSS Spectra are obtained for objects with r$<$17.77 as well as for luminous red galaxies described in Eisenstein et al (2001).

While the studies discussed above have provided large amounts of information on the optically bright counterparts to 1.4GHz sources, one requires deeper surveys to identify the bulk of the radio sources. 
Optical counterparts to sources in the Phoenix Deep survey (Hopkins et al. 2003) have been studied by Sullivan et al (2004). They investigate photometric redshifts for sources in a region with varying sensitivity -- the theoretical rms noise ranges from 10 $\mu $Jy to 0.36mJy. They have $UBVRI$ imaging for about 1 deg$^{2}$ but only have K-band observations for a $15^\prime\times 15^\prime$ region in which they could find only a handful of sources with S$>1$mJy. Since IR observations provide important constraints for photometric redshift estimation at higher redshifts, little information is provided on the redshift distribution of the brighter sources. Similarly, in the work by Ciliegi et al. (2005) which investigates optical counterparts to 80 $\mu $Jy sources, deep K-band data is only available in 165 arcmin$^{2}$ which allows potentially reliable photometric redshift estimation for a small number of sources and they do not address this explicitly.

  A complete subsample of the Leiden Berkeley Deep Survey (LBDS, Windhorst, van Heerde \& Katgert 1984) has been studied with deep optical and IR observations. The LBDS Hercules subsample consists of 72 sources observed at
$1.4$ GHz  with flux-density $S_{1.4}\ge 1$mJy in a 1.2$\;$deg$^{2}$ region of
Hercules. They imaged sources in {\it{G,R,I}} and {\it{K}} bands, identifying 69 of the sources in at least two bands. They also obtained spectra of 47 of the sources (Waddington et al. 2000). Photometric redshifts were estimated for the rest of sample using the spectral synthesis code of Jimenez et al. (1998). The data was used to examine the
evolution of the radio luminosity function and its high redshift cut-off
(Waddington et al. 2001). This study has allowed the redshift distribution and nature of 1.4GHz sources with S$>$1mJy to be studied fairly well but the area of sky is small and may not represent a fair sample. The Combined EIS-NVSS Survey of Radio Sources (CENSORS) survey (Best et al. 2003, Brookes et al. 2006) matches sources in the ESO Imaging Survey to 150 sources in the NRAO VLA Sky Survey (Condon et al. 1998) and is complete to 7.8mJy. Complete spectroscopic data from this survey will provide further information on the nature and redshift distribution of faint radio sources. 

Ultimately, we would like to study the evolution of clustering using measurements in different redshift slices and since it is very difficult to obtain spectra for a large number of optically faint sources, it is useful to investigate the use of photometric redshifts more extensively. Templates in the photometric redshift code used by Waddington et al. (2001) do not explicitly include light from AGN activity and it is useful to consider this (Brodwin et al. 2006).

An interesting application of the optical identifications of radio sources involves using the K-z relation to probe the nature of radio galaxy hosts. Lilly and Longhair (1984) showed that for 3CRR galaxies, (sources with flux-densities greater than 10\,Jy at 178MHz), there is a tight correlation between the K-band magnitudes of host galaxies and the redshift of the galaxy. The simplest interpretation of these results is that radio galaxies formed at high-redshift and have been evolving passively since then (see also Jarvis et al. 2001). Best et al. (1998) argued that this interpretation could not be correct since there is evidence that radio galaxies at high redshift are found in richer environments than those at low redshift. Studies of fainter radio sources provide more information on the nature of radio-galaxy hosts, with results from the 6C catalogues indicating an offset in the K-z relation to fainter magnitudes. Some studies (eg. Eales et al. 1997) claimed the offset only applies to z$>$0.6 galaxies but Willott et al. (2003) attribute the apparent lack of offset at low-z to the small number of sources in the sample. Their results, as well as those of McLure et al. (2004), are consistent with the idea that the K-z relation for galaxies with brighter radio flux-densities (S$>$500\,Jy at 151MHz) can be understood using the correlation between stellar mass and black-hole mass (Magorrian et al. 1998) combined with a correlation between black hole mass and radio luminosity. Best et al (1998) and de Breuck et al. (2001) also use this idea in explaining their results. While the reasoning involving black hole mass correlations works well for brighter sources, in McLure et al.(2004), it appears to break down for the fainter radio sources in the TexOx-1000 survey (Hill \& Rawlings 2003) and further studies of the K-z relation for fainter radio sources would help in understanding this.

Observations of extragalactic radio sources also provide important tests for galaxy evolution theory, particularly in clusters of galaxies where AGN are believed to heat the intracluster medium (eg Croton et al. 2006).

In this paper, we discuss the properties and K-z relation of optical counterparts of FIRST radio sources found in the Bo\"{o}tes field of the NOAO Deep-Wide Field Survey (NDWFS). Many studies are being carried out in this region. Using FIRST matches to NDWFS sources, Wrobel et al. (2005) identified 55 candidates suitable for the study of parsec-scale properties of radio sources and McGreer et al. (2006) searched for high-redshift quasars, finding a z=6.1 quasar. De Vries et al. (2002) and Wilman et al. (2003) studied the properties of radio sources in the Bo\"{o}tes field detected in a Westerbork 1.4\,GHz survey and pointed out that optical and IR identifications were important for further interpretation of their results. New photometric data at Spitzer wavelengths (Brodwin et al. 2006) and X-ray wavelengths (Brand et al. 2006), as well as a spectroscopic campaign called the AGN and Galaxy Evolution Survey (Brown et al., 2006) in the Bo\"{o}tes region, will provide further information on the radio sources in the field. The NDWFS data covers $\sim 9$ deg$^{2}$ ($216.1^{\circ}<$ RA $\le 219^{\circ}$, $32^{\circ}<$ Dec $\le 36^{\circ}$ ) but we only deal with the $\sim 5 $deg$^{2}$ between  $34^{\circ}<$ Dec $\le 36^{\circ}$ which has K-band data readily available.

In section 2 and 3 we
describe briefly the FIRST survey and the NDWFS. In
section 4 we present the results from the optical and infrared identification and, in section 5, we discuss the magnitude distributions of the optical counterparts. Section 6 deals with the photometric redshift estimation, the resulting redshift distributions and the K-z relation. In section 7, we summarise our results.

\section{The FIRST Survey}

The FIRST Survey (Faint Images of the Radio Sky at Twenty centimetres) covers
a total of $9033$ deg$^{2}$ of the sky: $8422$ deg$^{2}$ in the North Galactic cap and 611 deg$^{2}$ in the South Galactic cap. This yields a catalogue of $811 117$ sources with a positional accuracy better than $1''$ at $90\,\%$ confidence and flux-density limit of $1.0$ mJy at $1.4$ GHz. We have used the $3$ April $2003$ version of the FIRST  catalogue which is derived from the $1993$ through $2002$ observations. This catalogue is available from the FIRST website\footnote{http://sundog.stsci.edu/first/catalogs/}.

\section{The NOAO Deep-Wide Field Survey Catalogue}

The NDWFS is a deep optical and near infrared {\it{(Bw R I J H K)}} imaging survey that covers two $9.3$ square degree fields designed to study the formation and the evolution of large scale structure (Jannuzi $\&$ Dey $1999$). The first semester field (Bo\"{o}tes field ) is located near the north Galactic pole, centred on RA $= 14^{h}\, 32^{m}\,05.571^{s}$, dec $= +34^{\circ}\,16^{'}\, 47.5^{''}$ (J$2000$) ) and the second semester (Cetus field) is roughly $30^{''}$ from the South Galactic pole centred on RA of $2^{h}\, 10^{m}$ and dec $= -4^{\circ}\,30^{'}$.\\

This paper utilises the NDWFS third data release (DR3) obtained for the 27
subfields that compromise the Bo\"{o}tes fields of the NOAO Deep Wide Field
Survey. We have used the merged {\it{BwRIK}} catalogue for the
Bo\"{o}tes field which is available in ASCII tables from the NOAO website\footnote{http://www.noao.edu/noao/noaodeep/}. The NDWFS object catalogues were generated using the
SExtractor 2.3.2 (Brown et al. 2003). This catalogue has been split by
declination range into four strips ($32^{\circ}<Dec\le33^{\circ}$, $33^{\circ}<Dec\le34^{\circ}$, $34^{\circ}<Dec\le35^{\circ}$, $35^{\circ}<Dec\le36^{\circ}$)
and band (BwRIK) into $16$ files (each strip observed in four bands: Bw, R, I
and K). The merged catalogue contains 2 SExtractor flags which provide
information about the catalogue data quality: 1) FLAG\_DUPLICATE =1 indicates
that object is repeated elsewhere in the catalogue; 2) FLAG\_SPLITMATCH=1 is used
for objects with uncertain matches (dubious matches or photometry).

\begin{figure}

    \resizebox{80mm}{!}{\includegraphics[angle=270]{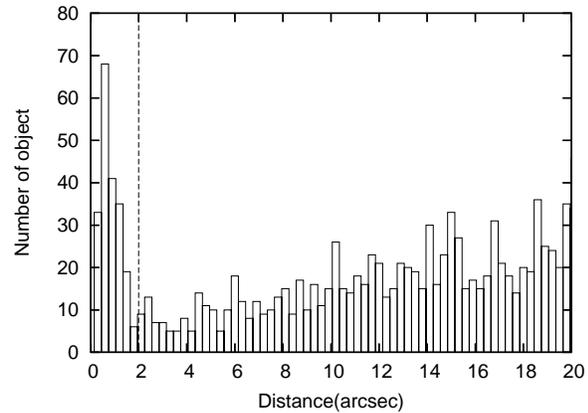}}
    \caption{Distribution of the radio-optical positional offsets between each
     FIRST radio source and the closest NDWFS object. The vertical dotted line
     indicates the distance cut-off ($2^{''}$) used to define the real
     radio-optical identifications.}
\end{figure}

\begin{figure}
  \begin{center}
      \resizebox{80mm}{!}{\includegraphics[angle=270]{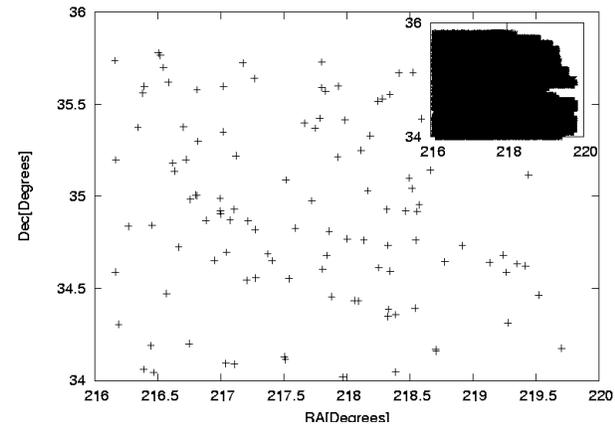}}

    \caption{Right ascension and declination distribution of the FIRST radio
      sources identified in all four bands in the NDWFS Bo\"{o}tes field. The inset shows the coverage of the K-band data.}
  \end{center}
\end{figure}

\section{Optical and Infrared Identification of FIRST radio sources}
\begin{table}
\begin{center}
\begin{tabular}{  c c c c c }

\hline
\hline
     &  $1^{st}$ Strip  & $2^{nd}$ Strip  & $3^{rd}$ Strip  & $4^{th}$ Strip   \\
\hline

 Bw      &       $397531$        &          $650832$     &         $790446$     &        $670230$     \\ 
 R       &       $339584$        &          $606205$     &         $605044$     &        $565157$     \\
 I       &       $320125$        &          $568260$     &         $652376$     &        $526129$     \\ 
 K       &       $---$           &          $3793$       &         $51122$     &       $40810$    \\ 
\hline
\end{tabular}
\end{center}
\caption{The number of sources in the NDWFS catalogue observed in Bw, R, I and K. Each column gives the number of sources 
in a $1^\circ$ strip, the first strip being $32^\circ<\delta\le 33^\circ$, the second being $33^\circ<\delta\le 34^\circ$ etc. Only the third and fourth strips are considered in this work. }
\end{table}

\begin{table}
\begin{center}
  \begin{tabularx}{\linewidth}{p{0.5cm}cccl}

\hline
\hline
      & $3^{rd}$ Strip  & \multicolumn{1}{c}{$4^{th}$ Strip}  & \multicolumn{1}{c}{Both Strip}    \\
           \multicolumn{2}{r}{$34<dec\le 35$}  & \multicolumn{1}{c}{$35<dec\le 36$}  & \multicolumn{1}{c}{$34\le dec\le 36$}       \\
\hline
                                  \multicolumn{4}{c}{Sources detected in one band}    &              \\

 Bw  &    $63\,\%$        &          $79\,\%$     &         $69\,\%$        \\
 R   &    $64\,\%$        &          $85\,\%$     &         $72\,\%$       \\
 I   &    $67\,\%$        &          $86\,\%$     &         $74\,\%$      \\
 K   &    $37\,\%$        &          $46\,\%$     &         $40\,\%$      \\
\hline
\hline
                                     \multicolumn{4}{c}{Sources detected in two
bands}  &\\
\multicolumn{1}{l}{BwR}      &       $58\,\%$        &          $74\,\%$     &
       $63\,\%$        \\
\multicolumn{1}{l}{BwI}       &       $57\,\%$        &          $72\,\%$     &
        $63\,\%$      \\
 RI       &       $60\,\%$        &          $78\,\%$     &         $67\,\%$
   \\
IK        &       $35\,\%$        &          $44\,\%$     &         $38\,\%$
  \\
\hline
\hline
                                       \multicolumn{4}{c}{Sources detected in three bands} & \\
\multicolumn{1}{l}{BwRI}      &       $54\,\%$        &          $69\,\%$     &
        $60\,\%$        \\
\multicolumn{1}{l}{BwIK}      &       $32\,\%$        &          $39\,\%$     &
        $35\,\%$       \\
\multicolumn{1}{l}{RIK}       &       $34\,\%$        &          $43\,\%$     &
        $38\,\%$       \\
\hline
\hline
                         \multicolumn{4}{c}{Sources detected in four bands}  &\\
 \multicolumn{1}{c}{BwRIK}      &       $32\,\%$       &          $38\,\%$
&         $34\,\%$      \\
\hline
\hline
\end{tabularx}
\end{center}
\caption{Optical and near infrared identification results for the FIRST radio sources.}
\end{table}

There are several techniques which could be used to cross-correlate radio and
optical catalogues. One statistically robust method is the likelihood ratio
technique of Sutherland \& Saunders (1992). This method has been  used often in order to identify radio sources (e.g. Gonzalez-Solares et al 2004, Ciliegi et al
2005, Sullivan et al 2004, Afonso et al 2005). However if the positional accuracies for both radio and optical catalogues are very high, positional coincidence alone can be adequate. Sullivan et al. (2004) demonstrated that the two methods produced very similar catalogues of matched objects in their study. We therefore positionally matched the FIRST radio sources with the two last strips of the NDWFS catalogue which cover $\sim$5 deg$^{2}$.  We
adopt a search radius of 2 arcsec as a good compromise when inspecting Fig 1,
where the distance distribution of the radio--optical separation is
presented. We use the K-band to estimate the contamination rate since they are the sparsest images. The number of sources observed in the $\sim$5 deg$^{2}$ of interest, in the K band is $9.2\,\times 10^{4}$ (see table 1), giving a surface density, $\rho$, of $1.4 \times 10^{-3}$ sources/arcsec$^{2}$. The number of FIRST sources in the region, $N_{F}$, is 514. Thus, with a search radius of $r_{s}$, the number of random coincidences is given by 
\begin{eqnarray}
N_{r}=N_{f}\times 2\pi\times r_{s}^{2}\times \rho \approx 18.
\end{eqnarray}
This gives a contamination rate of $3.5\%$. As a comparison, Magliocchetti $\&$ Maddox (2002) used a
2 arcsec cutoff for the APM-FIRST matches corresponding to $5\%$ random
matches and Prandoni et al. (2001) used a 3 arcsec cutoff for the ATESP-EIS (Australia Telescope ESO Slice Project and ESO Imaging Survey (EIS)) with
$2\%$ contamination rate.


We removed sources with uncertain matches in the merged catalogue (FLAG\_SPLITMATCH=1). We also used the SExtractor star-galaxy classifier to investigate subsamples of the matches: we considered a source to be extended if the stellarity parameter was $<0.5$. 

Figure 2 shows the position of FIRST sources identified in all four bands of the NDWFS  Bo\"{o}tes field.  

Table 1 shows the number of NDWFS sources in each band in each strip. Table 2 shows the results of the optical and IR identifications for the third and fourth strips in one band, two bands, three bands and four bands. The areas covered by the third and fourth strips are approximately equal yet there are 198 FIRST sources in the third strip and 316 FIRST sources in the fourth strip. The third strip also contains $\sim$15\% more sources in all bands. Interestingly, the identification rate is higher in the fourth strip where the source density is lower. This may be explained by the presence of large-scale structure in the region of the third strip, where many of the galaxies detected in the radio are too faint to detect optically - either because they are obscured by dust or are too distant - but systematics in the surveys should also be considered.

\section{Magnitude distribution}

\begin{figure*}
\begin{center}
\includegraphics{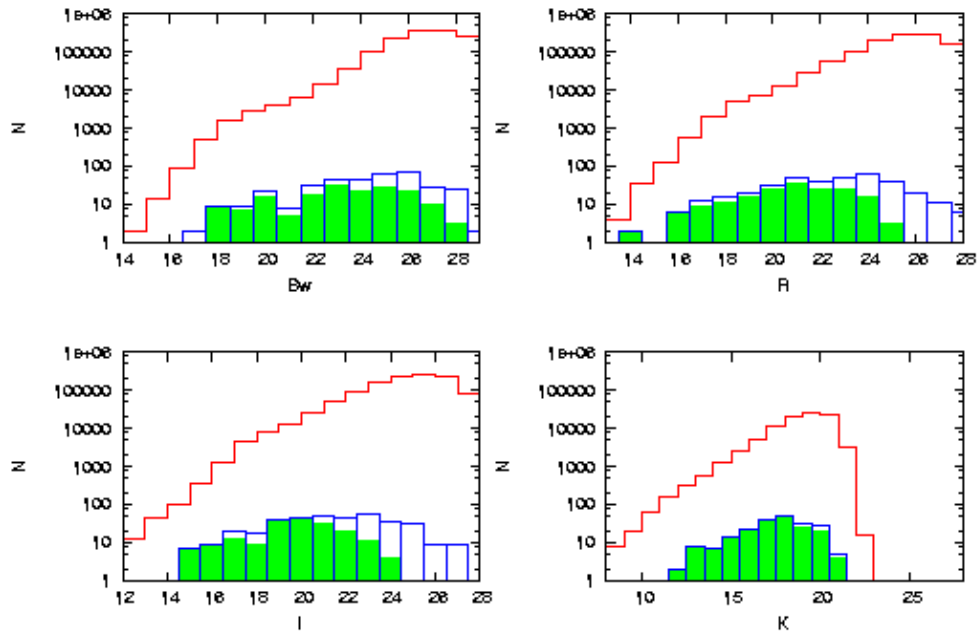}
    \caption{Magnitude distribution in the Bw, R, I and K band for, from top to      bottom: all sources in the NDWFS, the optical counterparts of the FIRST radio sources identified in the one band, the optical counterparts of the FIRST sources which have identifications in all four bands.}
\end{center}
\end{figure*}

In Figure 3, we compare magnitude distributions of sources identified in all four bands. Moving from top to bottom, the histograms shown are distributions for (i) all NDWFS sources in the region in the given band, (ii) the FIRST sources identified in the given band and (iii) the FIRST sources identified in all four bands. Comparing the first and second histograms, one notes that the
number of FIRST galaxies per magnitude interval does not 
continue to increase down to the magnitude limit, but rather turns over at Bw $\sim\,25$, r$\sim \,24$, I $\sim \,23$ and K$\sim \,18$. The turnover is $\sim 0.5$ magnitudes brighter than the magnitude limit for Bw, about a $\sim1.5$ magnitudes brighter for R and $\sim 2.5$mag brighter for I and K, highlighting the redness of the population of sources associated with mJy radio sources. The K-band is clearly shallower than the other bands.

\begin{figure*}
  \begin{center}
    \begin{tabular}{cc}
      \resizebox{60mm}{!}{\includegraphics[angle=270]{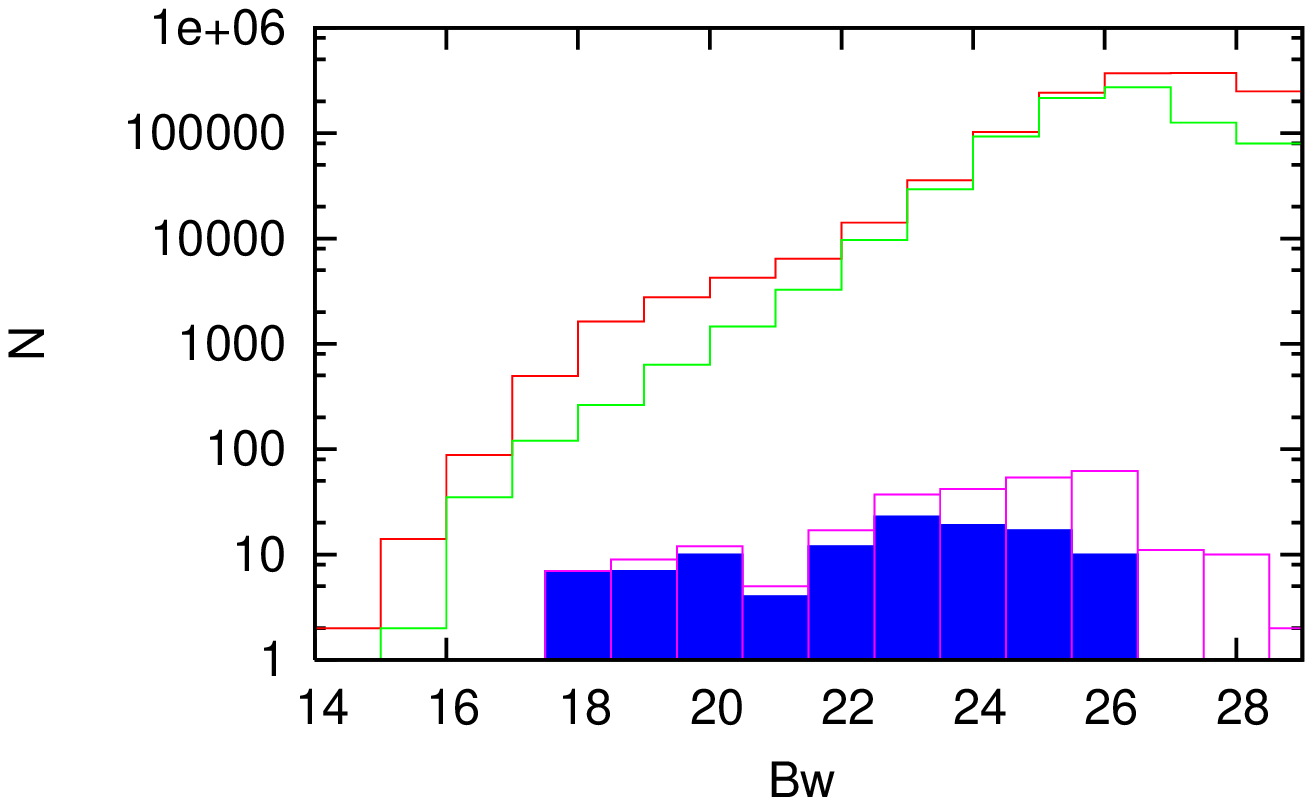}} &
      \resizebox{60mm}{!}{\includegraphics[angle=270]{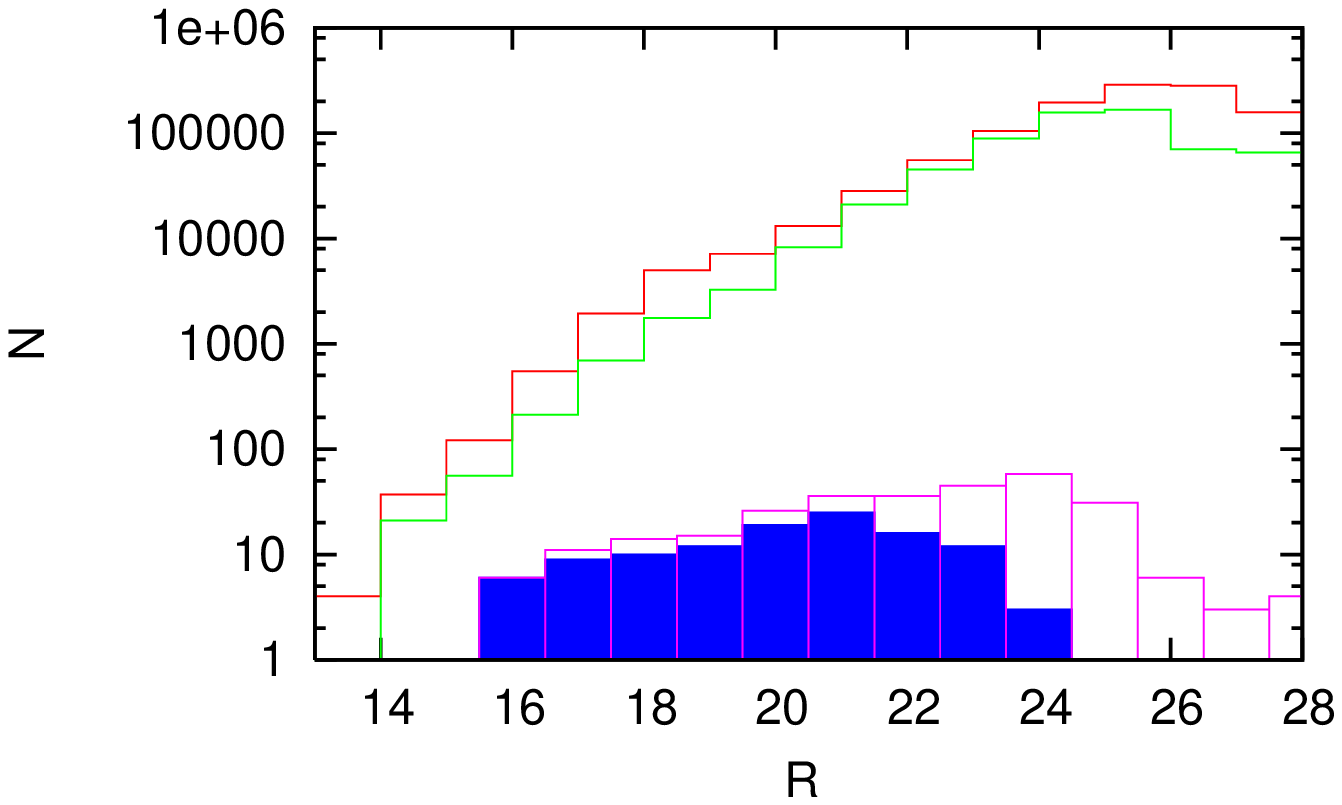}} \\
      \resizebox{60mm}{!}{\includegraphics[angle=270]{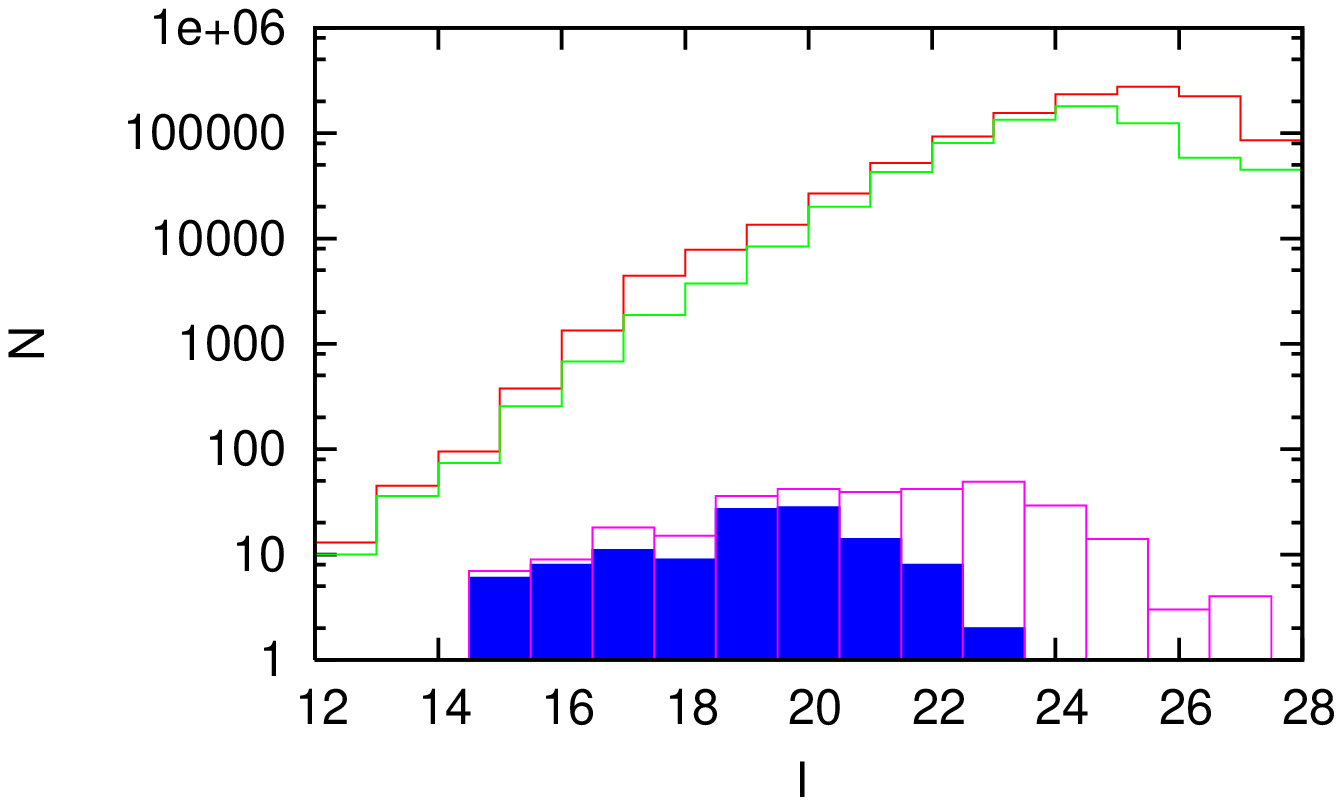}} &
      \resizebox{60mm}{!}{\includegraphics[angle=270]{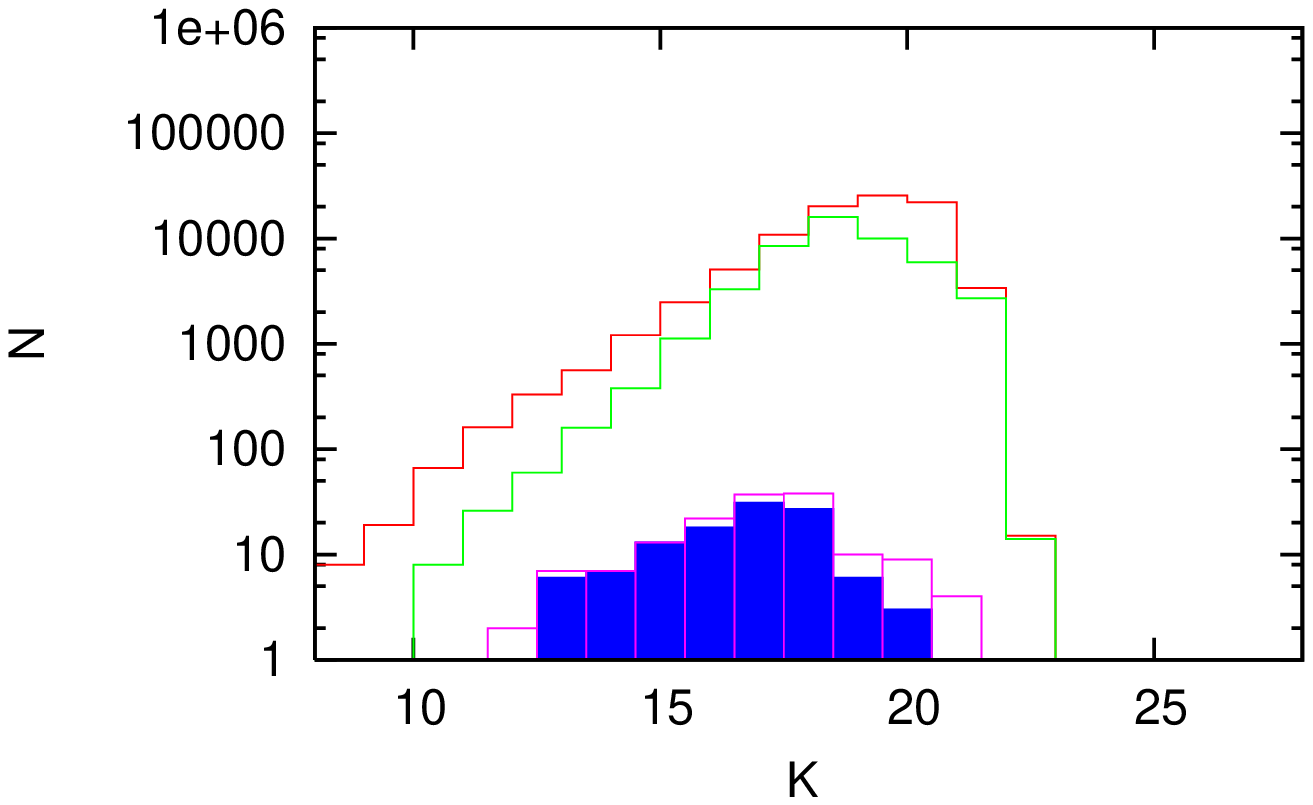}} \\
    \end{tabular}
    \caption{Magnitude distributions in the Bw, R, I and K bands for, from top to bottom, all sources in the NDWFS, sources in the NDWFS which are extended (have stellarity index $<0.5$), the extended optical counterparts of the FIRST radio sources identified in the one band, the extended optical counterparts of the FIRST sources which have identifications in all four bands.}
  \end{center}
\end{figure*}

In Figure 4, we investigate sources which are resolved in the NDWFS (have a SExtractor stellarity parameter of less than 0.5). Moving from top to bottom, the histograms show the distributions for (i) all NDWFS sources in the region in the given band, (ii) for all the NDWFS sources in the region with galaxy identifications in the given band, (iii) the FIRST sources identified in the given band and (iv) the FIRST sources identified in all four bands. Comparing the top two histograms, one sees that fainter sources are less likely to be characterised as galaxies, presumably because it is difficult to resolve more distant and intrinsically fainter sources, otherwise the trends are similar to those seem in Figure 3. 

It is interesting to note that in Waddington et al. (2001) 54 out of sources 72 (75\%) have K-band matches with K$<$19.4 while only 41\% of sourcs are idenitfied here. This supports the claim that studying areas larger than $\sim$1 deg$^{2}$ is required to get a fair sample of mJy sources and is perhaps to be expected when one considers that radio sources are clustered.

\begin{figure*}
  \begin{center}
    \begin{tabular}{cc}
      \resizebox{60mm}{!}{\includegraphics[angle=0]{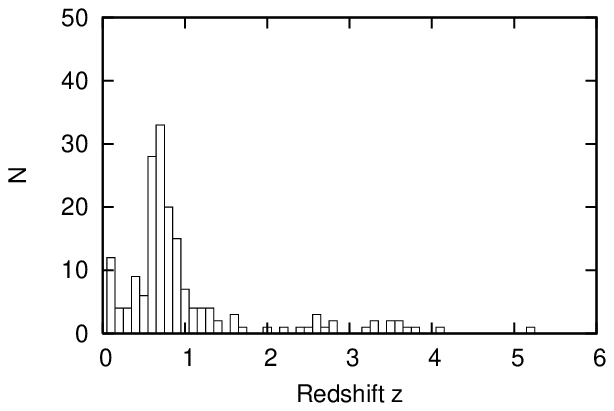}} &
      \resizebox{60mm}{!}{\includegraphics[angle=0]{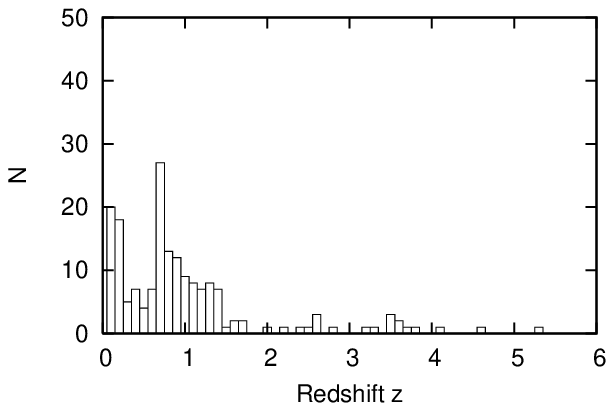}} \\
      \resizebox{60mm}{!}{\includegraphics[angle=0]{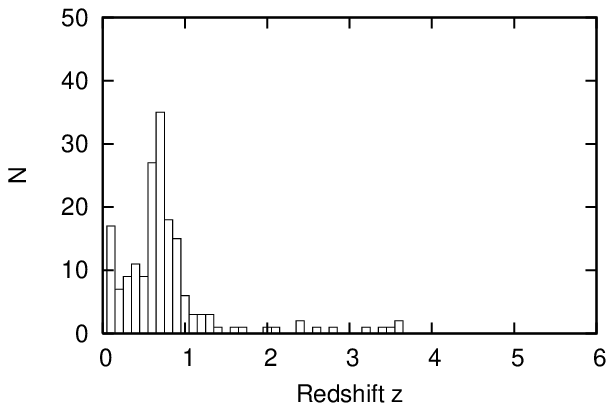}} &
      \resizebox{60mm}{!}{\includegraphics[angle=0]{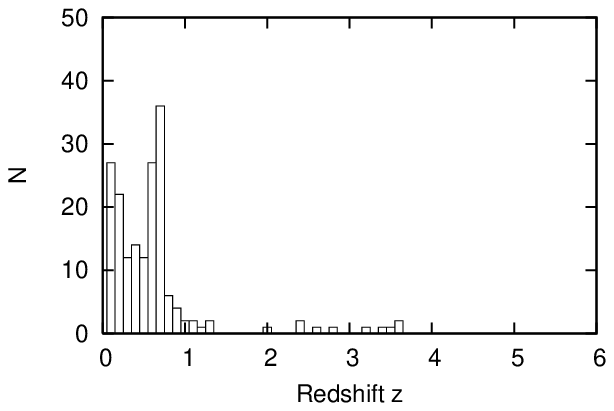}} \\
    \end{tabular}
    \caption{ Photometric redshift histograms for sources with counterparts in the four bands, as calculated by \textit{hyperz}. Top left: Quasar templates and burst templates included, M$>$-27. Top right: Quasar templates included, no burst templates, M$>$-27. Bottom left: No quasar templates, burst templates included, M$>$-24. Bottom Right: No quasar templates, no burst templates, M$>$-24.}
    \label{test4}
  \end{center}
\end{figure*}

\section{Photometric redshift distribution}

We estimated photometric redshifts for radio sources identified in all the
four bands: Bw, R, I and K, using the \textit{hyperz} code developed by
Bolzonella, Miralles \& Pello (2000). The code
takes a series of galaxy spectrum templates, varies the redshift and amount of
reddening and calculates photometric measurements that would be made through selected filters. These are matched to an input catalogue of photometric measurements using a standard $\chi^{2}$ minimisation procedure. We used GISSEL 98 galaxy
templates which are the 1998 update of the spectral synthesis models described by
Bruzual $\&$ Charlot (1993). All the models have solar metallicity and Miller
$\&$ Scalo (1979) initial mass function. Each model corresponds to a different
 present-day spectrophotometric galaxy type. Galaxies are represented 
by an exponentially decaying star formation rate with a
 time scale of 1\,Gyr for elliptical galaxies, increasing to 30\,Gyr for Sd galaxies. Irregular galaxies are represented by a constant star formation rate. The code also offers `delta-burst' templates corresponding to a single burst of star formation of a specific age.  

We adopted the reddening law of Calzetti et al. (2000). While there are large number of filters predefined in \textit{hyperz}, the
Bw filter is not included, we added the profile transmission of the filter
from the NDWFS website. We also investigated setting a maximum absolute magnitude for sources. The default limit is -29 while brightest cluster galaxies have a maximum V-magnitude of about -23.5. We limited the absolute magnitudes to -24, except when including QSO spectra when we allowed the absolute magnitude to be as large as -27.  

We investigated the addition of four QSO templates obtained from the LE PHARE website \footnote{http://www.lam.oamp.fr/arnouts/LE\_PHARE.html}. These template where selected for only 9/177 sources (5\%) and 5 of these were the highly reddened quasar template.  131 sources (74\%) where identified as 'bursts', 14 sources (8\%) where identified with ellipticals and 8 (5\%) with S0's leaving only 15 sources (8\%) identified with late-type galaxies. When the 'burst' option was removed, the reddest QSO template was selected as the best fit for 57/177 sources (32\%) and the elliptical template was selected for 76 (43\%) of sources. Probabilities for QSO matches and elliptical matches were significantly lower than those for burst options in most cases. 

When we limited the absolute magnitudes to -24 and removed the QSO templates. 138 of the sources were identified with 'bursts', 15 with ellipticals, 9 with S0's and 15 with later-type galaxies. When burst spectra and QSO spectra were not available, 138 sources (78\%) were identified with ellipticals. 

As in Bolzonella et al. (2000) we interpret the 'burst' spectra as elliptical galaxies with star formation on timescales shorter than 1 Gyr. When sources are identified with bursts, their ages range from 0.003 to 9.5 Gyr with a mean of 1.3 Gyr. When the burst option is removed and sources are identified with ellipticals, ages range from 0.01 to 12.5 Gyr with a mean age of 5.6 Gyr It is interesting to note that the age-to-formation implied for these 'ellipticals' is much longer than the ages implied for 'burst' spectra, suggesting that strong conclusions based on photometric age dating of ellipticals should be treated with caution. 

We also considered the subset of identifications which were resolved (SExtractor stellarity parameter $<0.5$). When QSO spectra are offered, 4\% are identified as such. When we limit the absolute magnitude to -24, 94/113 (83\%) of sources are then identified with the `delta burst' templates with a mean age of 1.6 Gyr. Four sources are identified with ellipticals, 5 with S0's and 10 with later type galaxies. The ages and split between different types of galaxies are similar to that obtained for all sources.

We tried to estimate photometric redshifts for sources that were only detected in three bands but found, by dropping one measurement for sources that were detected in all four bands, that estimates using three bands were not reliable.


Figure 5 shows the photometric redshift distribution for the FIRST radio
sources identified in the NDWFS Bo\"{o}tes field when various templates are included. The magnitude cut-off is -24 when quasars are not included and -27 when they are included. If we allowed quasar templates to be matched with a cutoff of -27 but require other templates to be matched with a -24 cutoff, the distribution looks very similar to that shown in the bottom left panel (since there are very few sources identified as quasars). One notes, in the top right panel, that matching to the red quasar template when bursts are not available, results in increase in sources in the 1$<$z$<$1.5 range. The burst option also tends to increase the number of lower redshift objects.

\begin{figure}
      \resizebox{80mm}{!}{\includegraphics[angle=0]{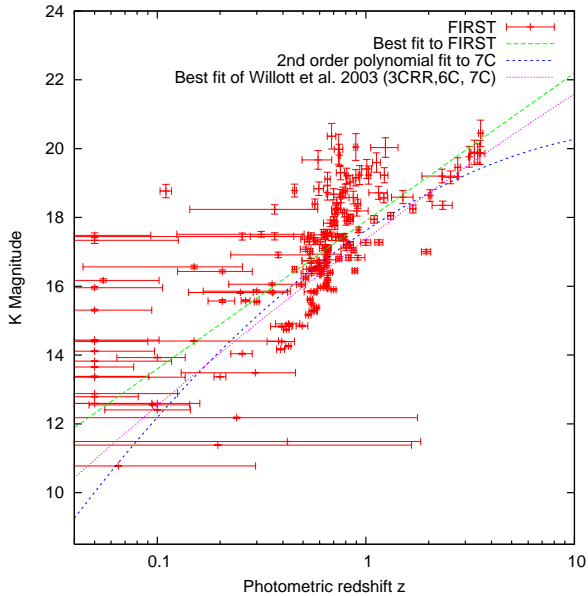}}
    \caption{The K-band magnitude versus redshift for FIRST-NDWFS matches. The upper line gives our fit to the FIRST data, the more curved line is the fit to 7C sources, similar to that obtained by Brookes et al (2006) for the CENSORS survey, and the third line is the fit to the 3CRR, 6C and 7C combination Willott et al. (2003).}
    \label{test4}
\end{figure}

\begin{figure*}
  \begin{center}
    \begin{tabular}{cc}
      \resizebox{60mm}{!}{\includegraphics[angle=270]{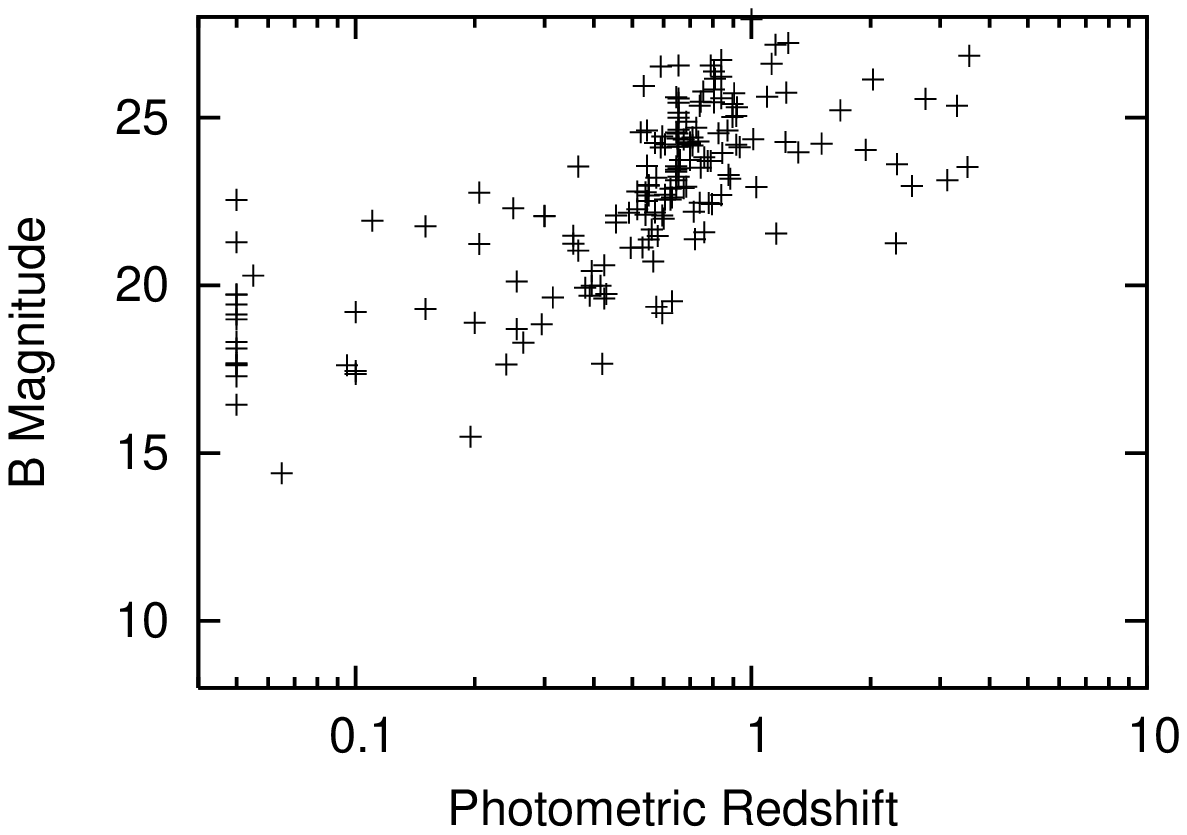}} &
      \resizebox{60mm}{!}{\includegraphics[angle=270]{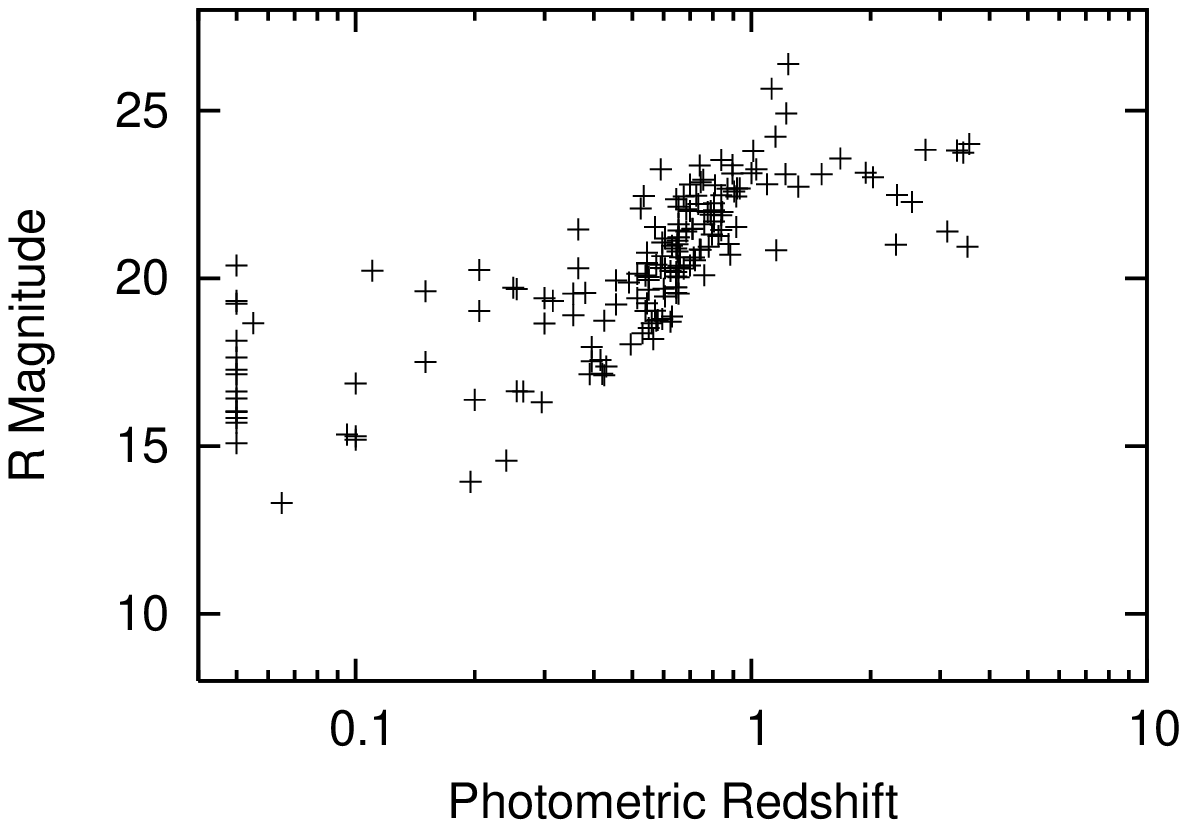}} \\
      \resizebox{60mm}{!}{\includegraphics[angle=270]{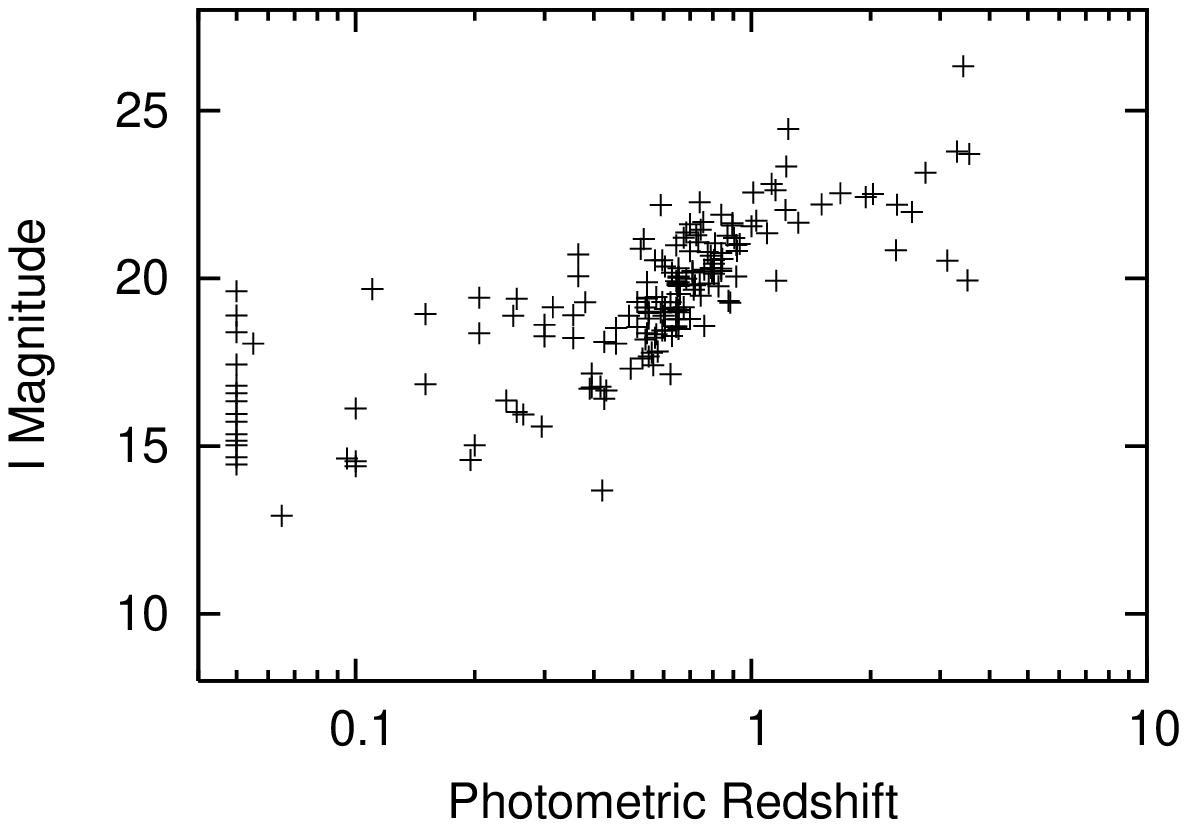}} &
      \resizebox{60mm}{!}{\includegraphics[angle=270]{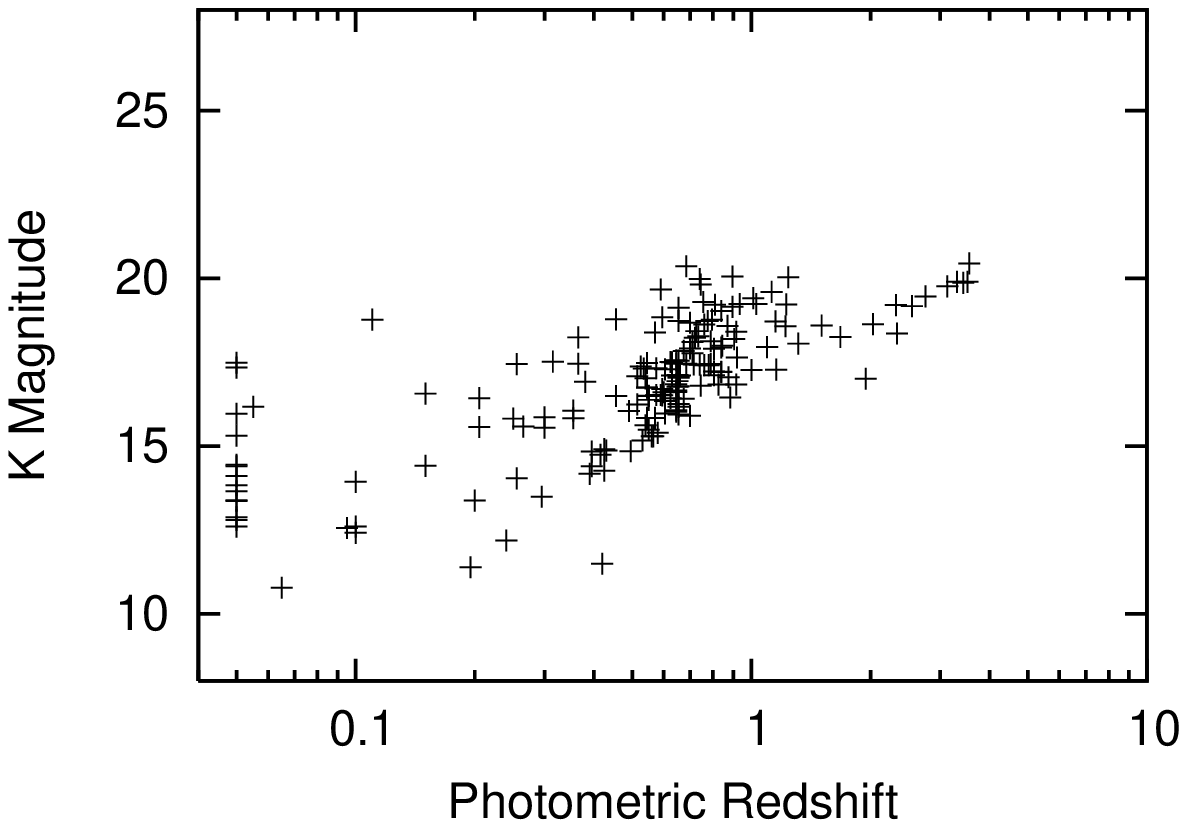}} \\
    \end{tabular}
    \caption{ Bw magnitude versus redshift (top left), R magnitude versus
      redshift (top right), I magnitude versus redshift (bottom right) and
      K magnitude versus redshift (bottom left) for the identified radio
      sources.}
  \end{center}
\end{figure*}

\section{The K-z relation}

In Figure 6 we plot the K-band magnitude of FIRST sources against photometric redshifts obtained for sources identified in all four bands of NDWFS (including unresolved sources). The redshifts correspond to those given in the bottom left panel of Figure 5 where a magnitude cutoff of -24 is used.  

The K-z relation for the combination of radio galaxies in 3CRR, 6C and 7C catalogues, given in Willott et al. (2003), is plotted in the figure (K=17.37$\,$+$\,$4.53$\,$log$_{10}$z$\,$-$\,$0.31$\,$(log$_{10}$z)$^2$). The fit for 7C alone, which Brookes et al (2006) claim is similar to the fit they obtain for the CENSORS survey, is also plotted. A fit to our data, given by K=(17.9$\pm $0.1)+(4.3$\pm $0.3)log$_{10}z$, is the line at faintest magnitudes. We have excluded the clump of sources at z=0.05 in the fit as these sources appear to be spurious identifications, possibly resulting from a lack of resolution in redshift during template matching. They are mostly identified with starforming galaxies and would not be expected to fit the K-z relation for radio galaxies. In the fit, we do not use the uncertainties on the redshifts generated by \textit{hyperz} as these are found to be unreliable when tested on samples where spectroscopic redshifts are available. Since FIRST sources are much fainter than 7C sources, which have flux-densities $S_{151MHz} >500$ mJy, and are selected at a much higher frequency, one might not expect the established K-z relation to hold. In addition, the uncertainties in the photometric redshift estimates are large and different aperture sizes for photometry could affect the results, so it is interesting that our relation is so similar to the other two. The fit for the FIRST sources is shifted slightly to fainter magnitudes which could support findings that brighter radio sources are associated with galaxies that are brighter in K, even for faint radio sources, but a more complete sample is required to investigate this further. We considered FIRST sources with radio flux-densities greater than 10mJy and did not find a better fit to the Willott et al. (2003) relation. 

In Figure 7, we plot the magnitude in each band against photometric redshift and find signs of a similar correlation in other bands, except with a larger scatter, particularly in I. The 1$\sigma$ variations are given by $\sigma_B= 2.65, \sigma_R=2.37,\sigma_I= 9.77,\sigma_K=1.96$. It is unclear why the I-band should have particularly large scatter. We note that the scatter in the K-band is significantly smaller at z$>$1 than at lower redshift, consistent with the idea that fainter radio surveys include fainter galaxies that are different from the typical galaxies associated with bright radio sources. Once again, a more complete sample and more reliable estimates of redshifts, which include more spectroscopic measurements, are required to investigate this further.

\section{Conclusion}
In this paper we have presented the optical and near infrared identifications
of $514$ radio sources in FIRST survey obtained by
matching objects in the NDWFS survey, over the region $216.1^{\circ}\le$ RA $\le 220^{\circ}$, $34^{\circ}\le$ DEC
$\le 36^{\circ}$. We identified 177 sources (35\%) in all 4 bands of NDWFS. In the I-band, 74\% of sources were identified. We found a surprisingly large difference between the number of sources in the upper declination strip and that in the lower strip, possibly explained by the presence of large-scale structure in the $34^{\circ}\le$ DEC $\le 35^{\circ}$ strip. Spectroscopy would be required to explore this further. Magnitude distributions demonstrated the red colours of the identified galaxies. The subset of sources which are extended (stellarity $<$0.5) in all bands (113/177) showed similar trends to the set including point sources but missing, perhaps, contributions from high-z objects which are difficult to resolve.    

We computed photometric redshifts for those sources with identifications in four bands using the code \textit{hyperz}.  The `burst' templates, corresponding to populations which formed over very short timescales, provided the best matches for the majority of sources. The ages implied for galaxies which match burst templates are significantly shorter than those implied if they are matched to elliptical templates. The four quasar templates we tried did not appear to fit the colours very well, although the dust-reddened QSO template was selected as the best match in many cases when the burst templates were not available. It appears that alternative approaches to photometric redshift estimation, such as those described in Brodwin et al. (2006) need to be explored.     

The K-z relation we obtained using photometric redshifts is similar to that obtained for brighter sources investigated by Willott et al. (2003) and Brookes et al. (2006) although there is much more scatter below z=1. Deeper K-band data and more use of spectroscopic redshifts is required before strong conclusions can be drawn. We plan to carry out a similar analysis for sources in the southern NDWFS field. which we can follow up spectroscopically using the Southern African Large Telescope.






\acknowledgements
    This work made use of images data products provided by the NOAO Deep Wide-Field Survey (Jannuzi and Dey $1999$), which is supported by the National Optical Astronomy Observatory (NOAO). NOAO is operated by AURA, Inc., under a cooperative agreement with the National Science Foundation.

\newpage

\end{document}